\documentclass[preprint,showpacs,showkeys,groupedaddress,superscriptaddress]{revtex4}
\usepackage{eucal}

\usepackage[usenames,dvipsnames]{xcolor}
\usepackage{hyperref}
\usepackage{graphicx}
\usepackage{amssymb, amsfonts}

\begin{document}

\title{Hamiltonian of mean force and a damped harmonic oscillator in an anisotropic medium }

\author{Marjan Jafari}
\footnote{M.Jafari\\Jafary. marjan@gmail.com, m.jafari@sci.ikiu.ac.ir}
\address{Department of Physics, Faculty of Science, Imam Khomeini International University, P.O.Box 34148 - 96818, Ghazvin, Iran}
\author{Fardin Kheirandish}
\address{Department of Physics, University of Kurdistan, P.O.Box 66177-15175, Sanandaj, Iran}

\begin{abstract}
The quantum dynamics of a damped harmonic oscillator is investigated in the presence of an anisotropic heat bath. The medium is modeled by a continuum of three dimensional harmonic oscillators and anisotropic coupling is treated by introducing tensor coupling functions. Starting from a classical Lagrangian, the total system is quantized in the framework of the canonical quantization. Following Fano technique, Hamiltonian of the system is diagonalized in terms of creation and annihilation operators that are linear combinations of the basic dynamical variables. Using the diagonalized Hamiltonian, the mean force internal energy, free energy and entropy of the damped oscillator are calculated.
\end{abstract}


\maketitle
\section{Introduction}\label{Introduction}
\noindent
Quantum damped harmonic oscillators appear in different branches of physics like condense matter \cite{10-6,12-6}, crystallin media \cite{8} and Quantum optics \cite{13-6,14-6}. Classical and quantum treatment of a damped harmonic oscillator has been extensively studied in \cite{1-2,2-2}.

The first attempts to describe the quantum dynamics of a damped harmonic oscillator was to find a Lagrangian or Hamiltonian from which desired equations of motion be derived. In this approaches, the effects of external fields where considered implicitly classically by considering Lagrangian or Hamiltonian of the system to be explicitly time-dependent. These schemes lead to some inconsistencies in fundamental principles of quantum mechanics like the violation of Heisenberg equations of motion. There are several approaches to investigate the quantum dynamics of a damped harmonic oscillator. In one of these approaches, known as the phenomenological approach \cite{21-6,23-6,21-m}, the fundamental ingredient is the fluctuation- dissipation theorem \cite{23-m}. In this approach, the dissipation is introduced to the system by adding some physically acceptable terms to the equations of motion. Another approach known as system plus reservoir approach, is based on Huttner-Barnett method where dissipation is introduced by modeling the reservoir by a continuum of harmonic oscillators interacting with the main system \cite{26-6,25-6}. The idea of a continuum set of harmonic oscillators, originally appeared in \cite{25-6}, has been extended and applied in electromagnetic field quantization in the presence of a medium and also in related topics such as macroscopic quantum electrodynamics and Casimir Physics \cite{8,amoo1,phil1,phil2,phil3}.

The thermodynamic equilibrium free energy of an open quantum system in contact with a thermal environment is equal to the difference between the free energy of the total system and free energy of the solely environment \cite{17-2}. The Hamiltonian of mean force is the effective Hamiltonian describing the Boltezmann-Gibbs equilibrium of the probability density of the open quantum system of interest \cite{188}.

In the present paper, we follow the second approach that is Huttner-Barnett approach and model the medium by a continuum of three dimensional harmonic oscillators. The anisotropic property is also taken into account by considering a tensorial or dyadic coupling function. Therefore, we start from a total Lagrangian and quantize the total system in the framework of canonical quantization. The Hamiltonian is diagonalized using Fano diagonalization technique \cite{fano}. The response or memory function, known also as susceptibility function, obeys Kramers-Kronig relations \cite{kk} and can be obtained in terms of dyadic coupling function and vice versa. The total system is assumed in a thermal equilibrium state and using the Hamiltonian of the mean force, the thermal-energy and free energy of the damped oscillator or main system are calculated in an anisotropic medium. The quantum dynamics of a damped harmonic oscillator in an isotropic medium has been discussed in \cite{phil2}.

The layout of the paper is as follows: In Sec. II, a classical Lagrangian for the total system is introduced, Langevin type equations of motion for the main system and reservoir oscillators are obtained. The response function and its connection to the dyadic coupling function is also discussed in this section. In Sec. III, The total system is quantized in the framework of canonical quantization approach and Hamiltonian is diagonalized using Fano diagonalization technique. In Sec. IV, thermal correlation functions, Hamiltonian of mean force, mean force thermal energy, free energy and entropy of the system are obtained. Finally, conclusions are given in Sec. V.
\section{Classical dynamics }\label{Classical dynamics}
\noindent
Following Huttner-Barnett approach, let us consider the total classical Lagrangian of the system plus reservoir as
\begin{equation}
L(t)=L_s+L_m+L_{int}.
\end{equation}
The first term $L_s$ is the Lagrangian of a unit mass harmonic oscillator or main system with displacement $\mathbf{q}$ and frequency $\omega_0$
\begin{equation}
L_s  = \frac{1}{2} \dot{\mathbf{q}}^2  - \frac{1}{2}\omega _0 ^2 \,\mathbf{q}^2,
\end{equation}
the second term, is the Lagrangian of a reservoir or heat bath consisting of a continuum of three dimensional harmonic oscillators with displacements $\mathbf{X}_{\omega}$ and frequencies $\omega \in [0,\infty )$
\begin{equation}
L _m= \frac{1}{2}\int\limits_0^\infty d\omega\,[\dot{\mathbf{X}}_\omega ^2 (t) - \omega ^2 \,\mathbf{X}_\omega ^2 (t)],
\end{equation}
and the last term is the interaction term
\begin{equation}
L_{{\mathop{\rm int}} }  = \sum\limits_{i,j} {\int\limits_0^\infty  {d\omega } } f_{ij} (\omega )\,q_i (t)\,X_{\omega, j} (t),
\end{equation}
where the main oscillator is coupled to the reservoir oscillators linearly and anisotropic property of the medium is considered by introducing dyadic or tensorial coupling functions $f_{ij}(\omega)$. For an isotropic medium we have $f(\omega)\delta_{ij}$. The classical equations of motion from Euler-Lagrange equations are
\begin{equation}\label{5}
\begin{array}{l}
 \frac{d}{{dt}}\left( {\frac{{\partial L}}{{\partial {\dot{q}} _i (t)}}} \right) - \frac{{\partial L}}{{\partial q_i (t)}} = 0,\,\,\,\,\,\,(i = 1,2,3) \\
 {\ddot{q}} _i (t) + \omega _0^2 \,q_i (t) = \int\limits_0^\infty  {d\omega } f_{ij} (\omega )X_{\omega, j} (t), \\
 \end{array}
\end{equation}
and
\begin{equation}\label{6}
\begin{array}{l}
 \frac{d}{dt}\left(\frac{\delta L}{\delta {\dot{X}} _{\omega, i} (t)}\right) - \frac{{\delta L}}{{\delta X_{\omega, i} (t)}} = 0,\,\,\,\,\,\,
 (i = 1,2,3) \\
{\ddot{X}} _{\omega, i} (t) + \omega ^2 X_{\omega, i} (t) = f_{ji} (\omega )\,q_j (t). \\
\end{array}
\end{equation}
In equations (\ref{5}, \ref{6}), the summation rule is applied over repeated indices. The formal solution of the equation (\ref{6}) is
\begin{equation}\label{8}
X_{\omega, i} (t) = {\dot{X}} _{\omega, i} (0)\frac{{\sin \omega t}}{\omega } + X_{\omega, i} (0)\cos \omega t + \int\limits_0^t {dt'}\,\frac{{\sin \omega (t - t')}}{\omega }f_{ji} (\omega )\,q_j (t'),
\end{equation}
the first term is the solution of the homogeneous equation and after quantization becomes a noise operator, the second term is the particular solution given by the Green's function of the harmonic oscillator
\begin{equation}\label{Green}
 G(t-t')=\Theta (t-t')\,\frac{{\sin \omega (t - t')}}{\omega },
\end{equation}
where $\Theta(t)$ is Heaviside step function. The dimensionless memory or response function $\chi_{ij} (t)$ is defined by
\begin{equation}\label{i0}
\chi _{ij} (t) =\frac{1}{\omega_0^2}\,\int\limits_0^\infty d\omega\,\frac{{\sin \omega t}}{\omega }\,f_{il} (\omega ) f_{jl} (\omega ).
\end{equation}
The real and imaginary parts of Fourier transform of response function $\tilde{\chi }_{ij}(\omega)$ satisfy Kramers-Kronig relations. From definition (\ref{i0}) and making use of inverse sine transform, one finds
\begin{equation}\label{Cgen}
  f_{ik} (\omega)\,f_{jk} (\omega)=\frac{2 \omega\omega_0^2}{\pi}\,\mbox[Im] [\tilde{\chi}_{ij} (\omega)].
\end{equation}
For convenience, from now we assume a symmetric coupling function ($f_{ij} (\omega)=f_{ji} (\omega)$), then
\begin{equation}\label{CC}
f_{ij} (\omega ) = \sqrt {\frac{2\omega\omega_0^2}{\pi}\mbox{Im}[\tilde{\chi} _{ij} (\omega )]}.
\end{equation}
So for a given susceptibility function, one can adjust coupling function according to (\ref{CC}). From (\ref{i0}), we find
\begin{equation}
 \omega_0^2\,\tilde{\chi} _{ij} (\omega ) = \mbox{P} \int\limits_0^\infty  {d\xi } \frac{{f_{il} (\xi )f_{lj} (\xi )}}{{\xi ^2  - \omega ^2 }} + i\pi \sum\limits_l {\frac{{f_{il} (\omega )f_{lj} (\omega )}}{{2\omega }}},
\end{equation}
where $\mbox{P}$ means principal value. Now by substituting (\ref{8}) into (\ref{5}), a classical Langevin equation is obtained for the main oscillator
\begin{equation}\label{LEq}
\ddot{q}_i (t) + \omega _0^2 \,q_i (t) -\omega_0^2\,\int\limits_0^t dt'\, \chi _{ij} (t - t')\,q_j (t') = \zeta _i^N (t),
\end{equation}
where
\begin{equation}\label{CNF}
  \zeta _i^N (t) = \int_0^\infty d\omega\, f_{ij} (\omega )\,\left(X_{\omega,j} (0)\,\cos(\omega t)+\dot{X}_{\omega,j} (0)\,\frac{\sin(\omega t)}{\omega}\right),
\end{equation}
is a classical noise force.
\section{Quantum dynamics}\label{Quantum dynamics}
\noindent To quantize the system plus reservoir in the framework of canonical quantization scheme, we need canonical conjugate variables corresponding to the dynamical variables $\mathbf{X}$ and $\mathbf{q}$. From total Lagrangian we have
\begin{eqnarray}
 \Pi _{\omega, i} (t) &=& \frac{\delta L}{\delta \dot{X} _{\omega, i}(t)} = \dot{X} _{\omega, i}(t), \nonumber\\
 p_i (t) &=& \frac{\partial L}{\partial \dot{q}_i } = \dot{q}_i (t).
\end{eqnarray}
The quantization is achieved by imposing equal-time commutation relations
\begin{eqnarray}
&& \left[ {\hat q_i (t),\hat p_j (t)} \right] = i\hbar \,\delta_{ij}, \\
&& \left[ \hat X_{\omega, i} (t),\hat \Pi_{\omega',j} (t) \right] = i\hbar\, \delta_{ij}\,\delta (\omega  - \omega '), \\
\end{eqnarray}
and all other equal-time commutation relations are zero. The Hamiltonian of the total system is obtained as
\begin{eqnarray}\label{H}
H &=& \frac{1}{2}\,\hat{\mathbf{p}}\cdot\hat{\mathbf{p}}+\frac{1}{2}\,\omega_0^2 \,\hat{\mathbf{q}}\cdot\hat{\mathbf{q}}+\frac{1}{2}\int\limits_0^\infty d\omega\, (\hat{\mathbf{\Pi}}_\omega\cdot\hat{\mathbf{\Pi}}_\omega+\omega ^2 \,\hat{\mathbf{X}}_\omega\cdot\hat{\mathbf{X}}_\omega)\nonumber\\
 &-& \frac{1}{2}\int\limits_0^\infty d\omega\,
(\hat{\mathbf{q}}\cdot \bar{\mathbf{f}}(\omega)\cdot\hat{\mathbf{X}}_\omega + \hat{\mathbf{X}}_\omega\cdot \bar{\mathbf{f}}(\omega)\cdot\hat{\mathbf{q}}),
\end{eqnarray}
where $\bar{\mathbf{f}}(\omega)$ is the dyadic coupling function. In Heisenberg picture, one finds the equations of motion as operator analogs of classical equations of motion (\ref{5}, \ref{6}). To have a better understanding of the quantum dynamics of the system and also subsystems, we try to diagonalize the Hamiltonian using Fano diagonalization technique. Therefore, we assume
\begin{equation}\label{58}
\hat{H} = \int\limits_0^\infty d\omega\, \hbar\omega\, \hat{\mathbf{C}}^{\dag} (\omega, t)\cdot\hat{\mathbf{C}} (\omega, t),
\end{equation}
that is the total Hamiltonian is assumed to be a continuum of uncoupled harmonic oscillators with creation and annihilation operators $\hat{\mathbf{C}}^{\dag}$ and $\hat{\mathbf{C}}$, respectively. These ladder operators satisfy bosonic commutation relations
\begin{equation}\label{66}
\left[ \mathbf{\hat C}(\omega ,t),\mathbf{\hat C}^\dag (\omega',t) \right] = \delta (\omega-\omega')\,\mathbb{I},\,\,\,\,\,\,\left[ \mathbf{\hat C}(\omega ,t),\mathbf{\hat C}(\omega ',t)\right] = 0,
\end{equation}
where $\mathbb{I}$ is a unit matrix. Generally, the annihilation operator $\hat{\mathbf{C}}$, is a linear combination of the original dynamical degrees of freedom
\begin{eqnarray}
\hat{\mathbf{C}} (\omega,t) = &-& \frac{i}{\hbar }\bigg[\bar{\mathbf{g}}^{*}_p (\omega )\cdot\hat{\mathbf{q}}(t) -
\bar{\mathbf{g}}^{*}_q (\omega )\cdot\hat{\mathbf{p}}(t)\nonumber \\
&+& \int\limits_0^\infty d\omega'\,\bar{\mathbf{g}}^{*}_{\Pi} (\omega ,\omega')\cdot\hat{\mathbf{X}}_{\omega'} (t)-
\bar{\mathbf{g}}^{*}_X (\omega,\omega')\cdot\hat{\mathbf{\Pi}}_{\omega'} (t)\bigg]. \nonumber\\
\end{eqnarray}
The following relations among coefficients can be determined from fundamental commutation relations
\begin{eqnarray}
&& g^*_{p,ij} (\omega)=i \omega\,g^*_{q,ij} (\omega),\\
&& g^*_{\Pi,ij} (\omega,\omega')=i \omega\,g^*_{X,ij} (\omega,\omega'),\\
&& -\omega\, g^*_{p,ij} (\omega)=-i \omega_0^2\,g^*_{q,ij} (\omega)+i\int_0^\infty d\omega'\,g^*_{X,ik} (\omega,\omega')\,f_{kj} (\omega'),\\
&& -i\omega\, g^*_{\Pi,ij} (\omega,\omega')={\omega'}^2\,g^*_{X,ij} (\omega,\omega')-g^*_{q,ik} (\omega)\,f_{kj} (\omega'),\\
&& \int_0^\infty d \omega''\,g_{\Pi,ik} (\omega,\omega'')\,g_{X,jk} (\omega',\omega'')=\frac{i\hbar}{2}\,\delta_{ij}\,\delta(\omega-\omega').
\end{eqnarray}
From these equations we find easily
\begin{equation}\label{gg1}
({\omega'}^2  - \omega ^2 )\,g_{X,jj'} (\omega,\omega') =g_{q,ji} (\omega )\,f_{ij'} (\omega '),
\end{equation}
\begin{equation}\label{gg2}
(\omega_0^2  - \omega ^2 )\,g_{q,jj'} (\omega ) = \int\limits_0^\infty  d\omega'\, g_{X,ji} (\omega,\omega')\,f_{ij'} (\omega').
\end{equation}
Equations (\ref{gg1}, \ref{gg2}) are similar to Eqs.(\ref{5}, \ref{6}) in frequency domain, so using (\ref{8}), the general solution of these equations can be written as
\begin{eqnarray}\label{73}
g_{X,ii'} (\omega ,\omega ') &=& h_{X,ii'} (\omega )\delta (\omega  - \omega ') \nonumber\\
&&+ \sum\limits_j {\frac{{f_{ji'} (\omega ')}}{{2\omega '}}} \left( {\frac{{\rm 1}}{{\omega ' - \omega  - i0^ +  }} + \frac{{\rm 1}}{{\omega ' + \omega }}} \right)g_{q,ij} (\omega ),
\end{eqnarray}
where $h_{X}(\omega)$ is an arbitrary tensorial function. The general solution for $g_{q}(\omega)$ is
\begin{equation}
g_{q,ii'} (\omega ) = h_{q,ii'} (\omega ) + \sum\limits_{j,l} {f_{ij} (\omega )h_{X,lj} (\omega )G_{li'} (\omega )}
\end{equation}
where the Green's function $G$ is defined by the inverse of the matrix $\Lambda_{li'}, (G=\Lambda^{-1})$
\begin{equation}\label{75}
\Lambda_{li'} (\omega ) = \left[(\omega_0 ^2  - \omega^2 )\delta _{li'}  - P\int\limits_0^\infty d\xi \,\frac{f_{i'j} (\omega )\,f_{jl} (\omega )}
{\xi ^2  - \omega ^2 } - i\pi\,\frac{f_{i'j} (\omega )f_{jl} (\omega )}{2\omega}\right],
\end{equation}
\begin{equation}\label{Gkapa}
G (\omega ) = \frac{-1}{\omega^2 \mathbb{I}-\omega _0^2 \,[\mathbb{I}-\tilde{\chi} (\omega )]},
\end{equation}
and $h_{q}$ is the solution of
\begin{eqnarray}\label{hq}
\left[(\omega_0 ^2  - \omega^2 )\delta _{li'} - P\int\limits_0^\infty d\xi \,\frac{f_{i'j} (\omega )\,f_{jl} (\omega )}
{\xi ^2  - \omega ^2 } - i\pi\,\frac{f_{i'j} (\omega )f_{jl} (\omega )}{2\omega}\right] h_{q,i'k} (\omega ) = 0.\nonumber\\
\end{eqnarray}
The explicit form of the function $h_X(\omega)$ can be determined from the commutation relations (\ref{66}) and (\ref{73}). One finds
\begin{equation}
\sum\limits_{ik} {2\omega h_{X,ik}^* (\omega )h_{X,kj} (\omega ) = \hbar } \delta _{ij},
\end{equation}
which has a simple solution
\begin{equation}\label{79}
h_{X,ij} (\omega ) = \left( {\frac{\hbar }{{2\omega }}} \right)^{\frac{1}{2}} \delta _{ij}.
\end{equation}
We can also show that the diagonalizing transformation requires the choice
\begin{equation}\label{82}
h_q(\omega)=0.
\end{equation}
The set of coefficients of the diagonalizing transformation is now determined by Eqs. (\ref{73}), (\ref{75}), (\ref{79}) and (\ref{82}). The canonical operators can also be expressed in terms of the annihilation and creation operators as
\begin{equation}
\hat q_i (\omega ) = 2\pi \left( {\frac{\hbar }{{2\omega }}} \right)^{\frac{{\rm 1}}{2}} \sum\limits_{l,j} {f_{il} (\omega )G_{lj} (\omega ) \hat C_j (\omega )}  = \frac{i}{\omega }\hat p_i (\omega ),
\end{equation}
\begin{eqnarray}\label{nnn}
\hat X_{\omega,i} (\omega') &=& 2\pi \sqrt{\frac{\hbar}{2\omega}} \sum\limits_j {\delta (\omega  - \omega ')\,\delta _{ij} \,\hat C_j (\omega )}\nonumber\\
 && + \sum\limits_j {\frac{{f_{ij} (\omega )}}{{2\omega }}\left(\frac{{\rm 1}}{{\omega  - \omega ' - i0^ +  }} + \frac{{\rm 1}}{{\omega  + \omega }}\right)\,\hat q_j (\omega ')}  = \frac{i}{{\omega '}}\,\hat \Pi _{X,i} (\omega ').\nonumber\\
\end{eqnarray}
\section{Thermal correlation functions}
\noindent
Having diagonalized Hamiltonian, now we proceed and find the thermal equilibrium expectation values of the internal energy and free energy of the main system in the framework of Hamiltonian of mean force. In global thermal equilibrium, we have
\begin{eqnarray}\label{222}
\langle \hat C_i^{\dag} (\omega )\,\hat C_j (\omega')\rangle  &=&  N(\omega )\delta (\omega-\omega')\,\delta _{ij} ,\,\,\,\,
N(\omega)=\exp (\hbar\omega/K_B T)-1, \\
 \langle \hat C_i^{\dag}  (\omega )\,\hat C_j (\omega')\rangle  &=& 0,
\end{eqnarray}
Using Eq.(\ref{222}) and straightforward calculations, we find for the symmetric thermal position and momentum correlations
\begin{equation}\label{stp}
  \langle \hat q_i(t)\,\hat q_j(t')\rangle_S= \frac{\hbar}{\pi}\int\limits_0^\infty d\omega\,\cos[\omega (t - t')]\,\coth \left(\frac{\hbar\omega}{2K_B T}\right)\,G_{ij} (\omega),
\end{equation}
\begin{equation}\label{stm}
\langle \hat p_i(t)\,\hat p_j(t')\rangle_S  = \frac{\hbar}{\pi}\int\limits_0^\infty  d\omega\,\omega^2\,
\cos[\omega (t - t')]\,\coth \left(\frac{\hbar\omega}{2K_B T}\right)\,G_{ij} (\omega),
\end{equation}
where the symmetric correlation is defined by
\begin{equation}\label{symcorr}
  \langle \hat q_i(t)\,\hat q_j(t')\rangle_S=\frac{1}{2}\,\langle \hat q_i(t)\,\hat q_j(t') + \hat q_j(t')\,\hat q_i(t)\rangle.
\end{equation}
From (\ref{Gkapa}) we obtain
\begin{equation}
 \frac{\pi}{2\omega}\,f_{ik} (\omega) f_{kl} (\omega) G_{lm}^* (\omega) G_{mj} (\omega)=\mbox{Im} [G_{ij} (\omega)].
\end{equation}
Having the explicit forms of the fields we can now find the thermal expectation values of reservoir and interaction parts of Hamiltonian as
\begin{eqnarray}
 && \frac{1}{2}\int\limits_0^\infty d\omega \,\langle (\partial_t \hat X_\omega )^2  + \omega^2 \,\hat X_\omega^2 \rangle\nonumber\\
 && = \frac{\hbar \,\omega_0^2}{2\pi}\mbox{Im} \int\limits_0^\infty d\omega\, \coth \left(\frac{\hbar \omega}{2K_B T}\right)\,\sum\limits_{i,j}
 \frac{d[\omega \,\chi_{ij} (\omega )]}{d\omega}\, G_{ji} (\omega ),
\end{eqnarray}
and
\begin{eqnarray}
&& \int\limits_0^\infty d\omega\, f_{ij} (\omega )\langle \hat q_i (t)\,\hat X_{\omega j} (t) \rangle\nonumber\\
&& = \frac{\hbar \,\omega _0^2 }{2\pi}\mbox{Im} \int\limits_0^\infty d\omega\, \coth \left(\frac{\hbar \omega}{2K_B T} \right)\sum\limits_l \chi_{il} (\omega )\,G_{lj} (\omega ).
\end{eqnarray}
 To find the internal energy, free energy and entropy of the main oscillator in the presence of an anisotropic heat bath, we use Hamiltonian of mean force. So in the next subsection we briefly discuss the Hamiltonian of mean force.
\subsection{ Hamiltonian of mean force} \label{ Hamiltonian of mean force}
\noindent Consider the total Hamiltonian of system plus reservoir as
\begin{equation}
 \hat H=\hat H_S+\hat H_R +\hat H_I,
\end{equation}
where $\hat H_I$ is the interaction term and $H_s$ are $H_R$ are Hamiltonian of system and reservoir respectively. The density operator of the total system is defined by
\begin{equation}\label{density}
  \rho=\frac{e^{-\beta H}}{Z},
\end{equation}
where $Z=\mbox{tr} \{\exp (-\beta H)\}$ is the the total partition function. The reduced density operator is given by $\rho_s=\mbox{tr}_R \{\rho\}$ and when the interaction term is negligible the reduced density operator is given by
\begin{equation}\label{reduced}
  \rho_s=\frac{e^{-\beta H_S}}{Z_S},
\end{equation}
with $Z_s=\mbox{tr}_S \{\exp (-\beta H_S)\}$. When the coupling between the system and its environment is not negligible then the reduced density matrix can not be written as (\ref{reduced}) but generically can be written as
\begin{equation}\label{MFD}
  \rho_s=\frac{e^{-\beta H^*_S}}{Z^*_S},
\end{equation}
where the Hamiltonian of mean force or effective Hamiltonian is given by
\begin{equation}
\hat H_S^* =- \frac{1}{\beta}\,\ln \left( \frac{\mbox{tr}_R [e^{-\beta \hat H}]}{Z_R} \right),
 \end{equation}
where $\beta=1/K_B T$ and $Z_R  = \mbox{tr}_R [\exp (-\beta \hat H_R )]$ is the partition function of reservoir. The partition function $Z^*$ associated with the Hamiltonian of mean force $\hat H^*$ is defined by
\begin{equation}\label{part}
Z^* = \mbox{tr}_S [\exp (-\beta \hat H_S^* )] = \frac{Z}{{Z_R }}.
\end{equation}
From (\ref{part}) we define the free energy of mean force
\begin{equation}
F^*  =  -\frac{1}{\beta}\,\ln (Z^* )= U - U_R - T(S - S_R ),
\end{equation}
where $U$ and $S$ are internal energy and entropy of total system. Now we can define the internal energy of mean force as
\begin{equation}\label{internal}
U^* = U- U_R  = \langle {\hat H}\rangle_{\mbox{tot}} - Z_R^{-1} \mbox{tr}_R [\hat H_R e^{-\beta \hat H_R }].
\end{equation}
For a harmonic oscillator interacting with an anisotropic reservoir, we find the internal energy of mean force as
\begin{eqnarray}
&& U^* =\langle \hat H \rangle_S \nonumber\\
&& = \frac{\hbar}{2\pi} \int\limits_0^\infty d\omega\, \coth \left(\frac{\hbar \omega}{2K_B T} \right)\mbox{tr}\left[\mbox{Im}
\left\{\left(\omega_0^2 \left[\omega\,\frac{d\,\bar{\chi}}{d\omega} - \bar{\chi} + 1 \right] + \omega^2\right) \,\bar{G}\right\} \right],\nonumber\\
&&= \frac{\hbar}{2\pi}\,\int\limits_0^\infty d\omega\, \coth \left(\frac{\hbar \omega}{2K_B T} \right)\mbox{tr}\left[\mbox{Im}
\left\{\frac{\omega_0^2 \left[\omega\frac{d\,\bar{\chi}}{d\omega}- \bar{\chi}+ 1 \right] + \omega^2 }{\omega_0^2
[1-\bar{\chi}]-\omega^2}\right\}\right],
\end{eqnarray}
which differs from the alternative definition \cite{Gelin}
\begin{eqnarray}
&& U =\frac{1}{2} \langle \hat p\cdot\hat p + \omega _0^2 \,\hat q\cdot\hat q \rangle\nonumber\\
&& = \frac{\hbar}{2\pi}\int\limits_0^\infty d\omega\,
\coth \left(\frac{\hbar \omega}{2K_B T}\right)\mbox{tr}\left[\mbox{Im}\left(\frac{\omega^2 +\omega _0^2}{\omega_0^2 \left[ 1 - \bar{\chi} (\omega) \right] - \omega^2 } \right)\right].
\end{eqnarray}
\subsection{Free energy and entropy}
\noindent The free energy of mean force can be obtained from $U^*=-T^2 \,\partial_T (F^* /T)$, as
\begin{eqnarray}
F^* &=& \frac{K_B T}{\pi}\int\limits_0^\infty d\omega\, \ln \left[\sinh \left(\frac{\hbar \omega}{2K_B T} \right) \right]\,\mbox{tr}\left[\mbox{Im}
\left\{\frac{\omega_0^2 \left[\omega\frac{d\,\bar{\chi}}{d\omega}- \bar{\chi}+ 1 \right] + \omega^2 }{\omega_0^2
[1-\bar{\chi}]-\omega^2}\right\}\right]\nonumber\\
&+& K_B T\,\ln 2,
\end{eqnarray}
and using the standard thermodynamic relation, $S^*=-\partial_T F^*$, the entropy of mean force is obtained as
\begin{eqnarray}
S^* &=& \frac{K_B T}{\pi}\int\limits_0^\infty d\omega\, \left\{\frac{1}{T}\,\coth\left(\frac{\hbar \omega}{2 K_B T}\right)-\frac{2 K_B}{\hbar \omega}\,
\ln \left[\sinh\left(\frac{\hbar \omega}{2 K_B T}\right)\right] \right\}\nonumber\\
&\times & \mbox{tr}\left[\mbox{Im}
\left\{\frac{\omega_0^2 \left[\omega\frac{d\,\bar{\chi}}{d\omega}- \bar{\chi}+ 1 \right] + \omega^2 }{\omega_0^2
[1-\bar{\chi}]-\omega^2}\right\}\right]+ K_B T\,\ln 2.
\end{eqnarray}
\section{Conclusion}
\noindent The quantum dynamics of a damped harmonic oscillator in the presence of an anisotropic heat bath investigated in the framework of canonical quantization. The medium modeled by a continuum of three dimensional harmonic oscillators and anisotropic coupling was treated by introducing tensor coupling functions. Tensorial response or memory functions was defined and its connection to coupling tensor determined. Following Fano technique, Hamiltonian of the system was diagonalized in terms of the creation and annihilation operators that are linear combinations of the basic dynamical variables. Using the diagonalized Hamiltonian, the mean force internal energy, free energy and entropy of the damped oscillator in the presence of an anisotropic medium were calculated.
\section*{References}



\begin{thebibliography}{10}
\expandafter\ifx\csname natexlab\endcsname\relax\def\natexlab#1{#1}\fi
\expandafter\ifx\csname bibnamefont\endcsname\relax
 \def\bibnamefont#1{#1}\fi
\expandafter\ifx\csname bibfnamefont\endcsname\relax
  \def\bibfnamefont#1{#1}\fi
\expandafter\ifx\csname citenamefont\endcsname\relax
  \def\citenamefont#1{#1}\fi
\expandafter\ifx\csname url\endcsname\relax
  \def\url#1{\texttt{#1}}\fi
\expandafter\ifx\csname urlprefix\endcsname\relax\def\urlprefix{URL }\fi
\bibitem{10-6} M. Born, K. Huang, \emph{Dynamical Theory of Crystal Lattices}, (NewYork, Oxford Univ,
Press 1985).
\bibitem{12-6} Y. J. Yan and R. X. Xu, \emph{Quantum mechanics of Dissipative systems}, Annu. Rev. Phys. Chem. \textbf{56}, 187 (2005).
\bibitem{8} L G Suttorp, J. Phys. A: Math. Theor. \textbf{40}, 3697 (2007).
\bibitem{13-6} C. W. Gardiner, P. Zoller, \emph{Quantum Noise A Handbook of Markovian and Non-
Markovian Quantum Stochastic Methods with Applications to Quantum Optics}, (Springer, 2000).
\bibitem{14-6} L. Mandel and E. Wolf, \emph{Optical Coherence and Quantum Optics}, (Cambridge University Press, Cambridge, 1995).
\bibitem{1-2} H. P. Breuer, F. Petruccions, \emph{The theory of open quantum systems}, (Oxford University Press 2002).
\bibitem{2-2} L. D. Landau, E. M. Lifshitz, and L. P. Pitaevskii, \emph{Electrodynamics of Continuous Media}, 2nd ed. (Butterworth-Heinemann, Oxford, 1984).
\bibitem{21-6} M. Razavy, \emph{Classical and Quantum Dissipative systems}, (Imperial College press, 2005).
\bibitem{23-6} G. Nicolis, I. Prigogine, \emph{Self-organization in Non-Equilibirium system}, (New York, Wiely, 1977).
\bibitem{21-m} S. Scheel, L. Knoll, and D. G. Welsch, Phys. Rev. A. \textbf{58}, 700 (1998).
\bibitem{23-m} U. Weiss, \emph{Quantum dissipative system}, (Singapore, World Scientific , 4th edn.)
\bibitem{26-6} B. Huttner, J. Baumberg, and S. Barnett, Europhys. Lett. \textbf{16}, 177 (1991).
\bibitem{25-6} B. Huttner and S. M. Barnett, Phys. Rev. A \textbf{46}, 4306 (1992).
\bibitem{amoo1} E. Amooghorban, F. Kheirandish, Int. J. Theor. Phys. \textbf{53}, 2593 (2014).
\bibitem{phil1} T.G. Philbin, J. Anders, J. Phys. A: Math. Theor. \textbf{49}, 215303 (2016).
\bibitem{phil2} T G Philbin, New j. phys. \textbf{14}, 083043 (2012).
\bibitem{phil3} T G Philbin, New J. Phys. \textbf{13}, 063026 (2011).
\bibitem{17-2} M. Campisi, P. Talkner P and P. H\"{a}nggi, Phys. Rev. Lett. \textbf{102}, 210401 (2009).
\bibitem{188} S. Hilt, B. Thomas, and E. Lutz, Phys. Rev. E \textbf{84}, 031110 (2011).
\bibitem{fano} L. G. Suttorp and A. J. van Wonderen, Europhys. Lett. \textbf{67}, 766 (2004).
\bibitem{kk} S. Scheel, L. Knöll, and D.-G. Welsch, Phys. Rev. A \textbf{58}, 700  (1998).
\bibitem{Gelin} M. Gelin and M. Thoss, Phys. Rev. E \textbf{79} 051121 (2009).
\end{thebibliography}
\end{document}